# Component Prototypes towards a Low-Latency, Small-form-factor Optical Link for the ATLAS Liquid Argon Calorimeter Phase-I Trigger Upgrade

Binwei Deng, Mengxun He, Jinghong Chen, Datao Gong, Di Guo, Suen Hou, Xiaoting Li, Futian Liang, Chonghan Liu, Gang Liu, Ping-Kun Teng, Annie C Xiang, Tongye Xu,    You Yang, Jingbo Ye, Xiandong Zhao, Tiankuan Liu

*Abstract*–This paper presents several component prototypes towards a low-latency, small-form-factor optical link designed for the ATLAS Liquid Argon Calorimeter Phase-I trigger upgrade. A prototype of the custom-made dual-channel optical transmitter module, the Miniature optical Transmitter (MTx), with separate transmitter optical sub-assemblies (TOSAs) has been demonstrated at data rates up to 8 Gbps per channel. A Vertical-Cavity Surface-Emitting Laser (VCSEL) driver ASIC has been developed and is used in the current MTx prototypes. A serializer ASIC prototype, operating at up to 8 Gbps per channel, has been designed and tested. A low-latency, low-overhead encoder ASIC prototype has been designed and tested. The latency of the whole link, including the transmitter latency and the receiver latency but not the latency of the fiber, is estimated to be less than 57.9 ns. The size of the MTx is 45 mm × 15 mm × 6 mm.

## I. Introduction

Large Hadron Collider (LHC) detectors, including the ATLAS Liquid Argon Calorimeter (LAr), widely use optical links to transmit detector data from the front-end detectors to the back-end counting room and timing/trigger/control information in the reverse direction [1-2]. When the LHC is upgraded in Phase-I to a high luminosity, the LAr trigger system needs to be upgraded to select efficiently events from background and fakes. The trigger system upgrade presents several challenges in optical link design.

The architecture of the proposed ATLAS LAr trigger electronics for the Phase-I upgrade is depicted in Fig. 1 [3]. In the figure, the new and upgraded components are shown in think bordered blocks. New LAr Trigger Digitizer Boards (LTDBs) will be developed and installed in the available spare slots of the front-end crates. Each LTDB samples and digitizes up to 320-channel detector signals and transmits the digitized data to new LAr Digital Processing Blade (LDPB) modules through 40 optical links. The LDPB modules calculate energies in real-time and provide high-granularity and high-precision information to upgraded trigger processors, called Feature EXtractors (FEXs). The FEXs extract the trigger primitives and transmit them to the Topological processors, which combine the data from both the calorimeter and the muon trigger modules and generate Level-1 trigger.

Two types of optical links, whose components are highlighted in green, between the LTDB and the LDPB are shown in Fig. 1. The uplink carries detector data from the LTDB to the LDPB, while the downlink carries timing (clocks) and control signals from the LDPB to the LTDB. The function blocks of the optical links are redrawn and shown in Fig. 2. On the uplink transmitter side, the electrical signal is converted to an optical signal in an optical transmitter module, which consists of a laser diode and a laser diode driver. At a distance of no greater than 300 meters, a Vertical-Cavity Surface-Emitting Laser (VCSEL) is more cost effective than an edge-emitting laser. Multiple-channel parallel data are multiplexed in a serializer and transmitted through a single fiber. The parallel data must be encoded before they are multiplexed in an encoder in order to maintain the DC balance of the serial data, limit the consecutive identical digits, and provide a mechanism for the receiver recover the parallel data boundary. The serializer and the encoder are usually implemented in a single chip which is called the transmitter. On the uplink receiver side, the optical signal is converted to the electrical signal in an optical receiver, which consists of a photodiode and a trans-impedance amplifier (TIA). The serial data are demultiplexed into parallel data in a deserializer. The original data are then recovered in the decoder. The deserializer and the decoder are usually implemented in a

Manuscript received November 15, 2013. This work is supported by US-ATLAS R&D program for the upgrade of the LHC, the US Department of Energy Grant DE-FG02-04ER1299, the US Department of Energy Collider Detector Research and Development (CDRD) and the National Science Council in Taiwan.

Binwei Deng is with Hubei Polytechnic University, Huangshi, Hubei 435003, P. R. China and with Department of Physics, Southern Methodist University, Dallas, TX 75275, USA.

Mengxun He was Arizona State University at Tempe, Tempe, Arizona 85281, USA. He is now with University of Texas at Dallas, Richardson, TX 75080, USA.

Jinghong Chen and Yang You are with Department of Electrical Engineering, Southern Methodist University, Dallas, TX 75275, USA and with Department of Electrical and Computer Engineering, University of Arizona, Tucson, AZ 85721, USA.

Datao Gong, Chonghan Liu, Annie C Xiang, Jingbo Ye, Xiandong Zhao, and Tiankuan Liu are with Department of Physics, Southern Methodist University, Dallas, TX 75275, USA (Tiankuan Liu: telephone: 214-768-1293, e-mail: tliu@smu.edu).

Suen Hou and Ping-Kun Teng are with Institute of Physics, Academia Sinica, Nangang 11529, Taipei, Taiwan.

Xiaoting Li is with Central China Normal University, Wuhan, Hubei 430079, P. R. China and with Department of Physics, Southern Methodist University, Dallas, TX 75275, USA.

Di Guo and Futian Liang are with University of Science and Technology of China, Hefei, Anhui 230026, P.R. China. Di Guo is and Futian Liang was also with Department of Physics, Southern Methodist University, Dallas, TX 75275, USA.

Gang Liu is with Institute of High Energy Physics, Chinese Academy of Sciences, Beijing 100049, P. R. China and was with Department of Physics, Southern Methodist University, Dallas, TX 75275, USA.

Tongye Xu is with Shandong University, Ji'nan 250100, China, P. R. China and was with Department of Physics, Southern Methodist University, Dallas, TX 75275, USA.

single chip which is called the receiver. The function blocks of the downlink are the same as those of the uplink.

Radiation tolerance is the first challenge in the optical link design. The components mounted on the detector operate in a harsh radiation environment [4], whereas the components located in the counting room are not exposed to radiation. Therefore, most components on the transmitter side of the uplink and the receiver side of the downlink are custom-designed for radiation tolerance. The components on the receiver side of the uplink and the transmitter side of the downlink can be implemented with Commercial-Off-The-Shelf (COTS) components. For the downlink, the GigaBit Transceiver (GBTX) ASIC [5] and the Versatile optical Transceiver module (VTRx) [6] will be used in the ATLAS LAr Phase-I trigger upgrade. This paper focuses on the transmitter side of the uplink, though the receiver side of the uplink will be discussed briefly for completeness. The downlink, however, is beyond the scope of this paper.

Low latency is the second challenge in the optical data link design. Latency is an important parameter in the trigger system because the latency determines the size of the event data buffer, where the detector data are stored to wait for the Level-1 trigger signal. The event data buffer, which is implemented in the Switched-Capacitor Array (SCA) analog pipeline on the Front-End Boards, will be kept unchanged in the ATLAS LAr Phase-I trigger upgrade. Therefore, the latency of the new developed sub-detectors is required to be no greater than that of the existing sub-detectors. The latency of 150 ns, not including the time passing through the optical fiber, is assigned to the optical link in the ATLAS LAr Phase-I trigger upgrade. In order to achieve the required low latency, a transmitter Application Specific Integrated Circuit (ASIC) with a custom-designed encoder is being developed.

Form factor is the third challenge in the optical link design. For an optical link, the optical transmitter module is larger than the transmitter ASIC. In the ATLAS LAr Phase-I trigger upgrade, each LTDB uses 40 optical links to transmit the data off the detector. Due to the limited front-panel space, the optical transmitter must be mounted on the board under the existing mechanical constraints. Therefore, the optical transmitter module must be small enough to allow 20 dual-channel modules on each LTDB. Clearance between the cooling plates and the LTDB motherboard is 6 mm. To avoid cut-out in the LTDB motherboard that could compromise its mechanical integrity, a small-form-factor optical transmitter module with a height of no greater than 6 mm must be developed. The form factor is limited by the size of the Lucent Connector (LC) packaged Transmitter Optical Sub-Assemblies (TOSAs). Such a dual-channel optical transmitter module called MTx is being developed.

The prototypes for a low-latency, small-form-factor optical link are primarily designed for the ATLAS LAr Phase-I trigger upgrade, but can potentially be used in other LHC upgrades with similar requirements.

The remainder of the paper is organized as follows: Section II describes the design and test results of the MTx. The design of a radiation-tolerant laser diode driver ASIC which is used in MTx is discussed in Section III. The serializer ASIC is discussed in Section IV. Section V discusses the encoder. The receiver implemented in an FPGA is discussed in Section VI. Section VII summarizes the paper.

## II. SMALL-FORM-FACTOR OPTICAL TRANSMITTER

MTx is a custom-made, two-channel, small-form-factor optical transmitter module based on VCSEL TOSAs. MTx adopts the design concept of tight integration of the transmitter and optical transmitter as the Small-Form-factor Versatile Transceiver (SF-VTRx) [6]. For both SF-VTRx and MTx, the transmitter is located on the motherboard underneath the optical transmitter module, which can be mounted at any position on the motherboard. SF-VTRx is specified with the height of 7 mm and the maximum data rate of 5 Gbps. MTx uses a different fiber interface in order to achieve a form factor of no greater than the 6 mm limited by the TOSAs. MTx uses different VCSEL drivers from SF-VTRx to achieve a data rate higher than 5 Gbps which is limited by the laser drivers used in SF-VTRx. MTx is compared to other optical transmitters in Table I.

TABLE I
THE COMPARISON OF MTX AND OTHER OPTICAL TRANSMITTER MODULES

|  | Size (mm) | Maximum data rate (Gbps) |
|---|---|---|
| SFP+ | 48.7 × 14.5 × 9.7 [a] | 10 |
| VTRx/VTTx | 45 × 14.5 × 10 | 5 |
| SF-VTRx | 45 × 15 × 7 | 5 |
| MTx | 45 × 15 × 6 | 8 |

[a] The size includes the cage.

The CAD drawing of the MTx module is shown in Fig. 3. MTx is composed of an optical connector **latch**, a **module** Printed Circuit Board (**PCB**), two **TOSAs** with flexible cables, a VCSEL driver ASIC called **LOCld1** which will be discussed in Section III, and an **electrical connector**. The CAD drawing of the latch is shown in Fig. 4. Since the height of a regular LC connector is greater than 6 mm, the outer case of the LC connector was discarded and the **fiber** with only the standard **flange**, the **ferrule** and the **spring** was order. The latch consists of two pieces. **Piece 1** has two notches to hold the TOSAs, as well as three **pins** and one **screw hole** to fix the TOSAs on the module PCB. **Piece 2** fixes the two fibers to Piece 1 with two **hooks**. The TOSA packages guarantee the alignment of the fibers with the VCSELs. The springs keep the fibers in a good contact with the TOSAs. The latch used in the prototype is produced by a 3-D printer. The final production of the latch will be injection molded with polyetherimide, which has been tested to be radiation tolerant [7]. The TOSAs used in the prototype have also been tested to be radiation tolerant [8]. The electrical connector is a Samtec 0.50-mm Razor Beam high speed hermaphroditic terminal/socket strip connector (Part Number LSHM-120-2.5-L-DV-A-N-T-R-C) that matches the height of TOSAs. In order to reduce the module height, the module PCB has two rectangular holes where part of the TOSA bodies can sink in the module PCB. All of the components are installed on one side of the module PCB.

An MTx prototype has been demonstrated. A picture of an MTx module is shown in Fig. 5 and a picture of an MTx module plugged in a carrier board is shown in Fig. 6. An eye diagram of the MTx prototype at 10 Gbps is shown in Fig. 7. The eye diagram passes the 10-Gbps fiber channel eye mask, indicating that the design goal is achieved. The input signal is a $2^7$-1 Pseudo-Random Binary Sequence (PRBS) with differential amplitude of 200 mV (peak-to-peak). The bias current is set at 6 mA and the modulation current is set at 6.4 mA. The average optical power is about -0.87 dBm. The power consumption of the MTx prototype is about 400 mW.

When one laser driver and one TOSA of MTx are replaced by a radiation tolerant TIA-embedded receiver optical sub-assembly (ROSA) [9], MTX can be changed into an optical transceiver. A Miniature optical Transceiver (MTRx) with the same form factor as MTx has been demonstrated and can be used for the downlink.

### III. THE LASER DRIVER ASIC

A radiation-tolerant VCSEL driver is needed in the MTx. A single-channel 8-Gbps VCSEL driver prototype, called LOCld1, has been developed and tested [10-11].

LOCld1 is designed and fabricated in a commercial 0.25-μm Silicon-on-Sapphire (SoS) CMOS technology. The SoS technology has been proven to provide good total ionizing dose (TID) tolerant characteristics [12]. The block diagram of LOCld1 is shown in Fig. 8. LOCld1 includes six pre-drive stages and one output stage with 50-Ω pull-up resistors. All stages are powered by 3.3 V supply. In order to achieve 8-Gbps operation, an active-inductor shunt peaking technique [13] is used in the pre-drive stages. In this design, the peaking strength can be adjusted in order to achieve optimal performance [14]. LOCld1 is AC-coupled with an individual VCSEL TOSA. Digital-to-analog-converters (DACs), an I²C slave module, and 16-bit internal registers are included. The modulation current, VCSEL bias current, and shunt-peaking strength are programmable via an I²C configuration interface. Considering that the SoS CMOS technology has a smaller single-event-upset (SEU) cross section than bulk CMOS technologies, no special SEU mitigation techniques in the design has been applied except for Triple Modular Redundancy (TMR) technique in the internal registers. LOCld1 is packaged in a 24-pin QFN package.

LOCld1 has been assembled in an MTx prototype and tested with a 200-mV (peak-to-peak) differential $2^7$-1 PRBS signal. The modulation current is programmable from 7.8 mA to 10.6 mA and the bias current is programmable from 2 mA to 14 mA. The eye diagram has been shown in Fig. 7.

Table 2 shows a brief comparison of LOCld1 and GBLD [15], a radiation-tolerant laser driver used in VTRx and SF_VTRx. GBLD is fabricated in a 130-nm CMOS technology with the target data rate of 5 Gbps. GBLD can provide up to 2 × 12 mA modulation current (there are two drivers that can be connected in parallel) and up to 43 mA bias current and drive both a VCSEL and an edge-emitting laser (EEL). LOCld1, on the other hand, is only designed to drive a VCSEL.

TABLE II
THE COMPARISON OF LOCLD1 AND GBLD

| Features | GBLD | LOCld1 |
|---|---|---|
| Technology | 130 nm CMOS | 0.25 μm SoS |
| Laser type | VCSEL/EEL | VCSEL |
| Design data rate (Gbps) | 5 | 8 |

### IV. THE SERIALIZER ASIC

Two serializer ASIC prototypes have been designed and tested. The first is a single-channel 5-Gbps serializer ASIC called LOCs1, which has been tested with a 200-MeV proton beam and proven to be suitable for the ATLAS LAr Phase-I upgrade [16]. The second is a two-channel serializer ASIC called LOCs2 [11], each channel operating at up to 8 Gbps. This section focuses on LOCs2.

LOCs2 is designed and fabricated in the same SoS CMOS technology as LOCld1. The block diagram of LOCs2 is shown in Fig. 9. LOCs2 is comprised of two 16:1 serializer channels. Each serializer channel operates at data rates of up to 8 Gbps. The input of each serializer channel is 16-bit parallel data in LVDS logic, and the output of each serializer channel is serial data in current mode logic (CML). Each serializer channel is composed of 4 stages of 2:1 multiplexers in a binary tree structure. The first stage of 2:1 multiplexer is implemented using static CMOS D-flip-flops (DFFs). The last three stages use CML DFFs in order to achieve a higher speed. Each serializer channel has a CML driver which is composed of five stages of CML differential amplifiers. An active-inductor shunt-peaking technique is used in the first four stages to increase the bandwidth. The last stage has 50-Ω pull-up resistors to match the 100-Ω differential output impedance. The two serializer channels share one LC-tank-based PLL, which provides clock signals to each serializer channel. The loop bandwidth of the PLL is programmable from 1.3 to 6.8 MHz for flexibility. Based on the results of LOCs1 [16], no special SEU mitigation techniques have been applied in the design.

LOCs2 is packaged in a 100-pin QFN package. Testing results show that LOCs2 works from 6.6 to 8.5 Gbps, which is limited by the tuning range of the LC-PLL. The eye diagram of LOCs2 at 8 Gbps is shown in Fig. 10. The power consumption of LOCs2 is 1.25 Watt. The random jitter of the PLL, measured through a clock signal generated by dividing the VCO output by a factor of eight, is less than 1 ps (RMS). The total jitter (peak-peak at the BER of $10^{-12}$) of the serializer serial data output is about 25 ps when the output of LOCs2 is a $2^7$-1 PRBS signal.

## V. THE ENCODER ASIC

The encoder, internally called LOCic, is designed to process the data before they are sent to the serializer. Although standard encodings exist in industry, yet after studying several such industrial encodings, we found that none meet the requirements of this project. As such, a custom encoding has been proposed and the encoder ASIC prototype has been designed and tested.

The frame definition of the LOCic is shown in Fig. 11. The input data of LOCic come from 8-channel Analog-to-Digital Converters (ADCs) sampling at the LHC bunch crossing clock (shown as the frame clock in the figure) of 40 MHz. Each channel of ADCs has a resolution of 12 bits and outputs in serial accompanying a serial data clock [17]. Some ADC implementation requires two extra bits per channel for calibration [18]. The digitized data and optional calibration data, shown as D0 - D13 in the figure, are user data. In LOCic, 16-bit frame control code (T0 - T15) is added at the end of the user data to form a data frame. The encoding efficiency is 87.5% and 85.7% with and without the calibration bits, respectively. The user data are scrambled to keep the DC balance.

The control code $T_0$–$T_7$ is an 8-bit cyclic redundant checking (CRC) code which is used to detect data transmission errors. The polynomial $P(x) = x^8+x^5+x^3+x^2+x^1+x^0$ is chosen because it is optimal in terms of Hamming distance with CRC size of 8 bits and user data size of 112 bits [19]. The control code $T_8$–$T_{11}$, "1010," serves as the frame boundary identifier and limits the length of consecutive identical digits (CIDs) to be no greater than the frame length. The remaining four bits ($T_{12}T_{13}T_{14}T_{15}$) are called the bunch cross identification (BCID) field and used on the receiver side to provide the BCID information to align the different channels during calibration. The field is formed from a PRBS and is a secondary frame boundary identifier because the field is predictable from the same fields in the previous frames.

The ASIC prototype has been fabricated in the same SoS CMOS technology as LOCld1 and LOCs2. The block diagram of the ASIC is shown in Fig. 12. The prototype has been tested in a laboratory environment. The latency of the ASIC encoder has been simulated and verified in the prototype. The FIFO takes 1-2 clock cycles of the 640-MHz clock. The latency varies after each power cycle due to the phase uncertainty of the internal 640-MHz clock which will be generated by dividing the high-speed serializer clock of 2.56 GHz by 4. The PRBS generator, the CRC generator, the scrambler and frame builder take one cycle of the 640-MHz clock. In total, the latency of the encoder is no greater than 6.25 ns, or 4 cycles of the 640-MHz clock.

A single ASIC, called LOCx2, which integrates two channels of encoders and serializers, is being developed for the ATLAS LAr Phase-I trigger upgrade. The data rate of each channel is determined to be 5.12 Gbps based on the selection of ADCs. The latency of the LOCx2 is estimated to be less than 10.9 ns. The power consumption of LOCx2 is estimated to be about 1 W.

Table III compares LOCx2 and GBTX [5], a radiation-tolerant transceiver developed for LHC upgrades. GBTX is fabricated in a 130-nm CMOS technology. It includes a transmitter channel and a receiver channel. The design data rate of GBTX is 4.8 Gbps. GBTX provides GBT and 8B/10B encodings, as well as an extra option of no encoding. The GBT encoding provides the forward error correction (FEC) capability with an efficiency of 70%. The 8B/10B encoding provides limited error detection and no error correction capability with an efficiency of 73%. The latency of GBTX is 212.5 ns in GBT mode and 237.5 ns in 8B/10B mode. The total power of GBTX is about 2.2 W.

TABLE III
THE COMPARISON OF LOCIC AND OTHER ENCODING

| Features | LOCx2 | GBTX |
|---|---|---|
| Technology | 0.25 µm SoS | 130 nm CMOS |
| Functions | 2 transmitters | 1 transmitter + 1 receiver |
| Encoding | LOCic | GBT, 8B/10B |
| Efficiency | 87.5%,[a] 85.7%[b] | 70%,[c] 73%[d] |
| Latency (ns)[e] | 57.9[f] | 212.5,[c] 237.5[d] |
| Data rate (Gbps) | 5.12 | 4.8 |
| Power (W) | 1.0[f] | 2.2 |

[a] With calibration bits
[b] Without calibration bits
[c] GBT mode
[d] 8B/10B mode
[e] The latency of the whole
[f] Estimated values

## VI. THE RECEIVER IMPLEMENTATION

The receiver, including a deserializer and a decoder, has been implemented in a Xilinx Kintex-7 FPGA. The block diagram of the implementation in Kintex-7 is shown in Fig. 13. With an input reference, the deserializer recovers a 320-MHz clock from the high-speed serial data stream for all other function blocks and converts the serial data stream into 16-bit parallel data. The synchronizer identifies the frame boundary. The data extractor retrieves the user data after the frame boundary is identified. The BCID generator recovers the 12-bit BCID information using the 4-bit PRBS fields in the current frame and in previous frames. The descrambler recovers the original user data. The CRC checker detects if the user data are transmitted correctly. Besides the user data, the decoder outputs a 12-bit BCID counter, a CRC flag, and a frame flag indicating whether the data are valid. Each receiver uses 1 gigabit transceiver, 365 registers, 522 lookup tables, and 183 slices in Kintex 7.

The latency of the receiver depends on the deserializer and the decoder. All unnecessary function blocks are bypassed and the operation clock frequency is raised to as high as possible to reduce the latency. The latencc of each function block of the decoder implemented in an FPGA can be conveniently measured by using the ChipScope Pro Analyzer tool. The latency of the synchronizer and the data extractor, which operate simultaneously, is 3 cycles (9.375 ns) of the 320-MHz clock. The latencies of the descrambler and the CRC checker are both one cycle of the 320-MHz clock (3.125 ns). The BCID generator takes two cycles (6.25 ns) of the 320-MHz clock, matching the timing of the CRC checker. The deserializer's latency, which was measured by using a high-speed real-time oscilloscope, ranges from 28.5 to 31.4 ns. Latency varies after each power cycle due to the phase uncertainty of the deserializer's recovered clock. The latency of the whole link, including the transmitter implemented with an ASIC and the receiver implemented in an FPGA, is estimated to be no greater than 57.9 ns. The latency variation can be absorbed when the data are latched with the LHC bunch crossing clock and sent to the following trigger system. In other words, the latency of the whole link is fixed in the scale of the bunch crossing clock cycle.

## VII. Conclusion

Several components towards a low latency, small-form-factor optical link designed for the ATLAS liquid argon calorimeter Phase-I trigger upgrade have been presented. The latency of the whole link, including the latencies of the transmitter and the receiver but not the latency of the optical fiber, is estimated to be no greater than 57.9 ns. The size of the MTx is 45 mm × 15 mm × 6 mm.

## Acknowledgment

We are grateful to Drs. Sandro Bonacini and Paulo Moreira of CERN for sharing the design of the I$^2$C slave and the LVDS receiver, as well as Francois Vasey and Csaba Soos of CERN and Janpu Hou of FOCI for reviewing the MTx design. We would like to thank Mrs. Jee Libres and Nicolas Moses of VLISP Technologies, Inc. for beneficial discussions.

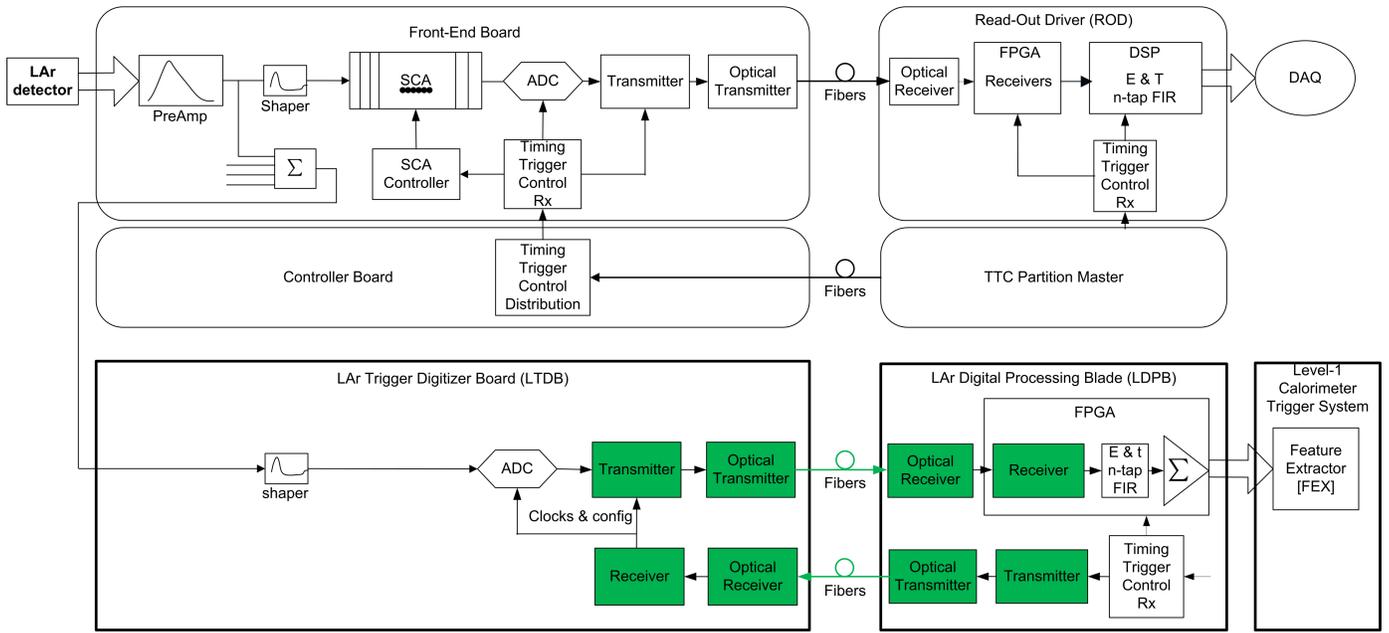

Fig. 1. Schematic block diagram of the Phase-I upgrade LAr trigger readout architecture.

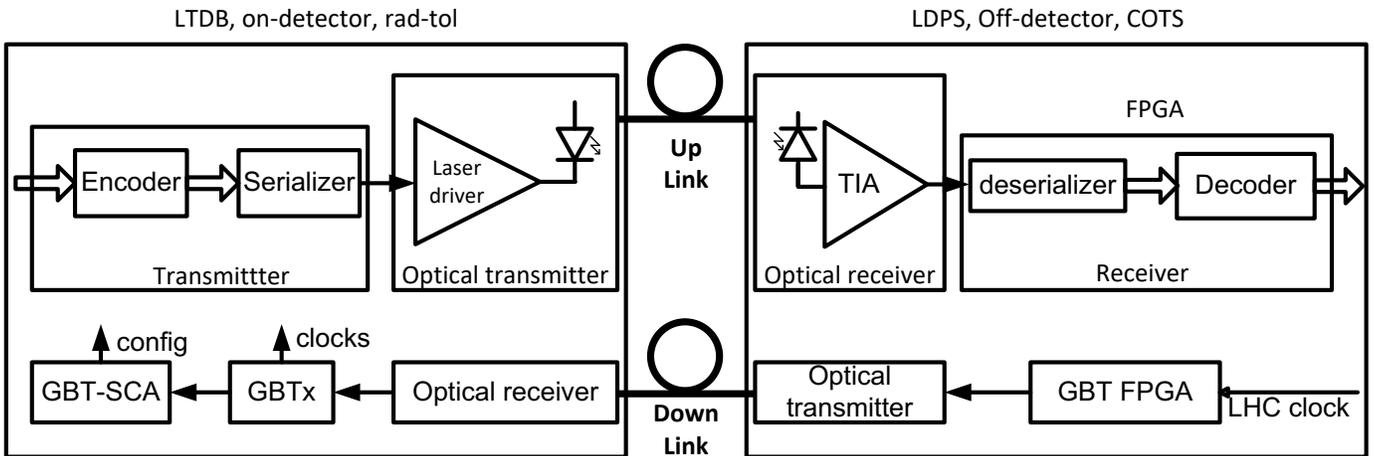

Fig. 2. Block diagram of a typical optical link.

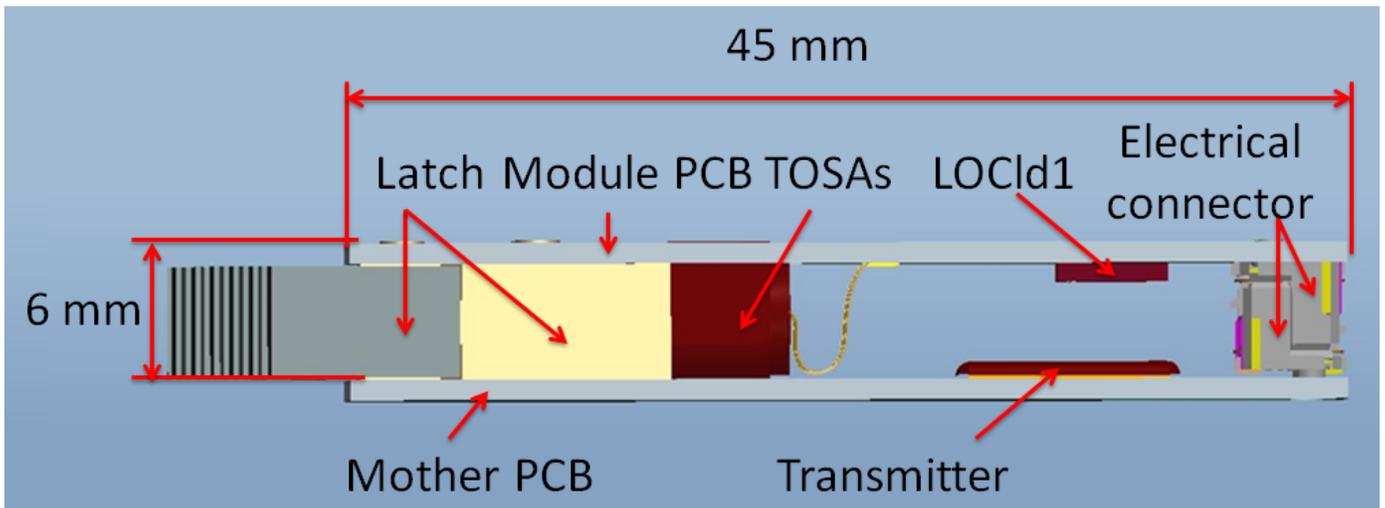
Fig. 3. CAD drawing of the MTx module

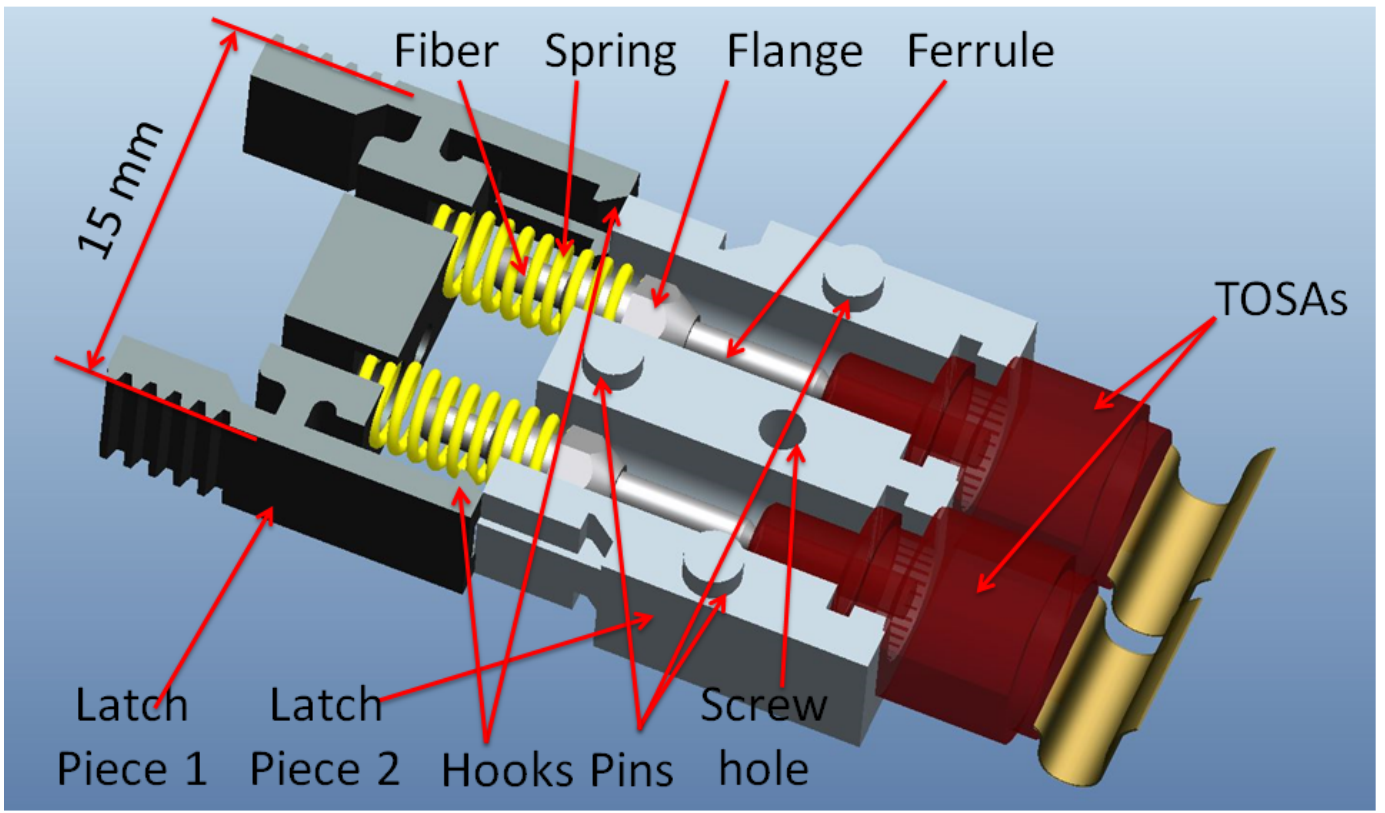
Fig. 4. CAD drawing of the optical connector latch

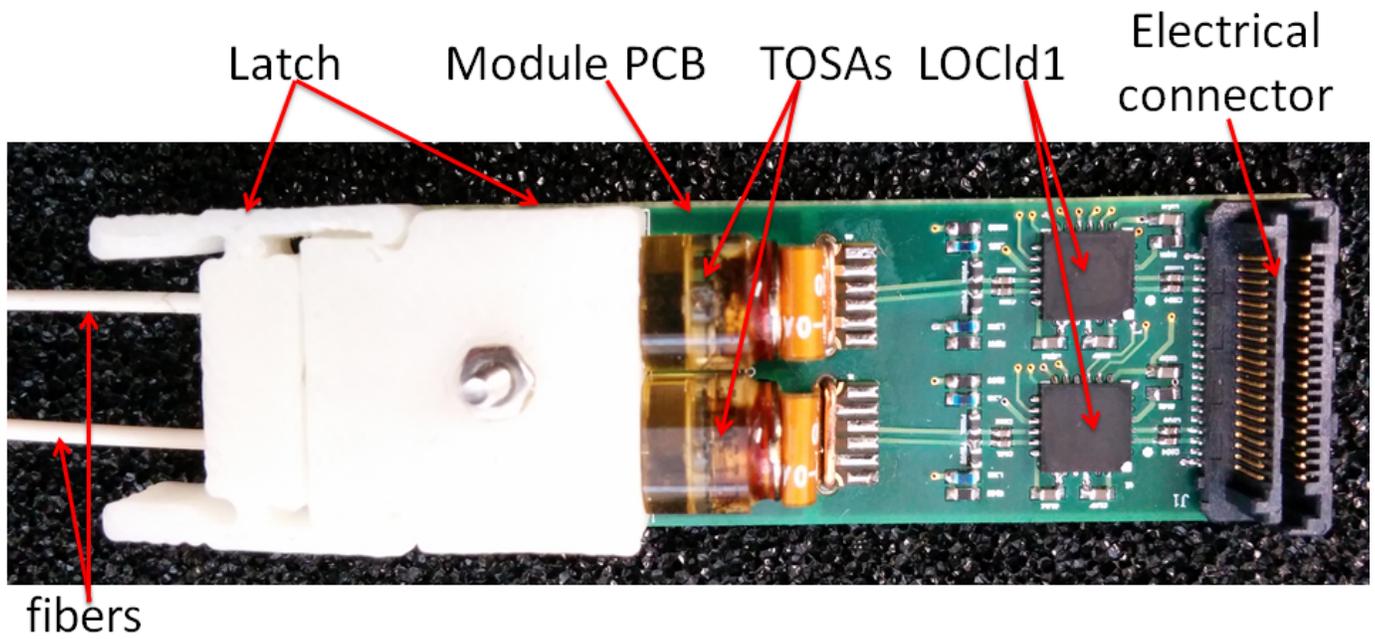

Fig. 5. An MTx module

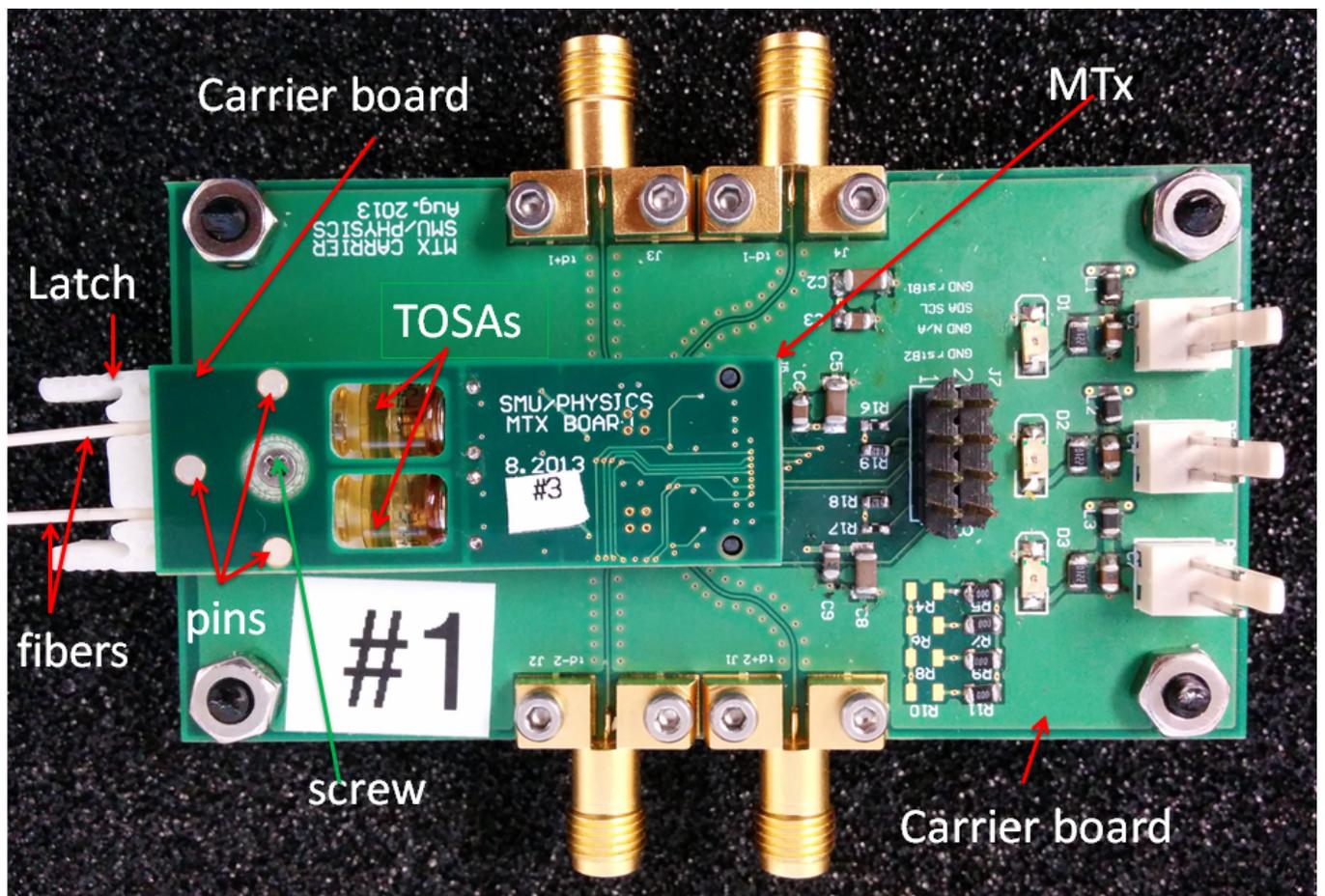

Fig. 6. An MTx module plugged into a carrier board.

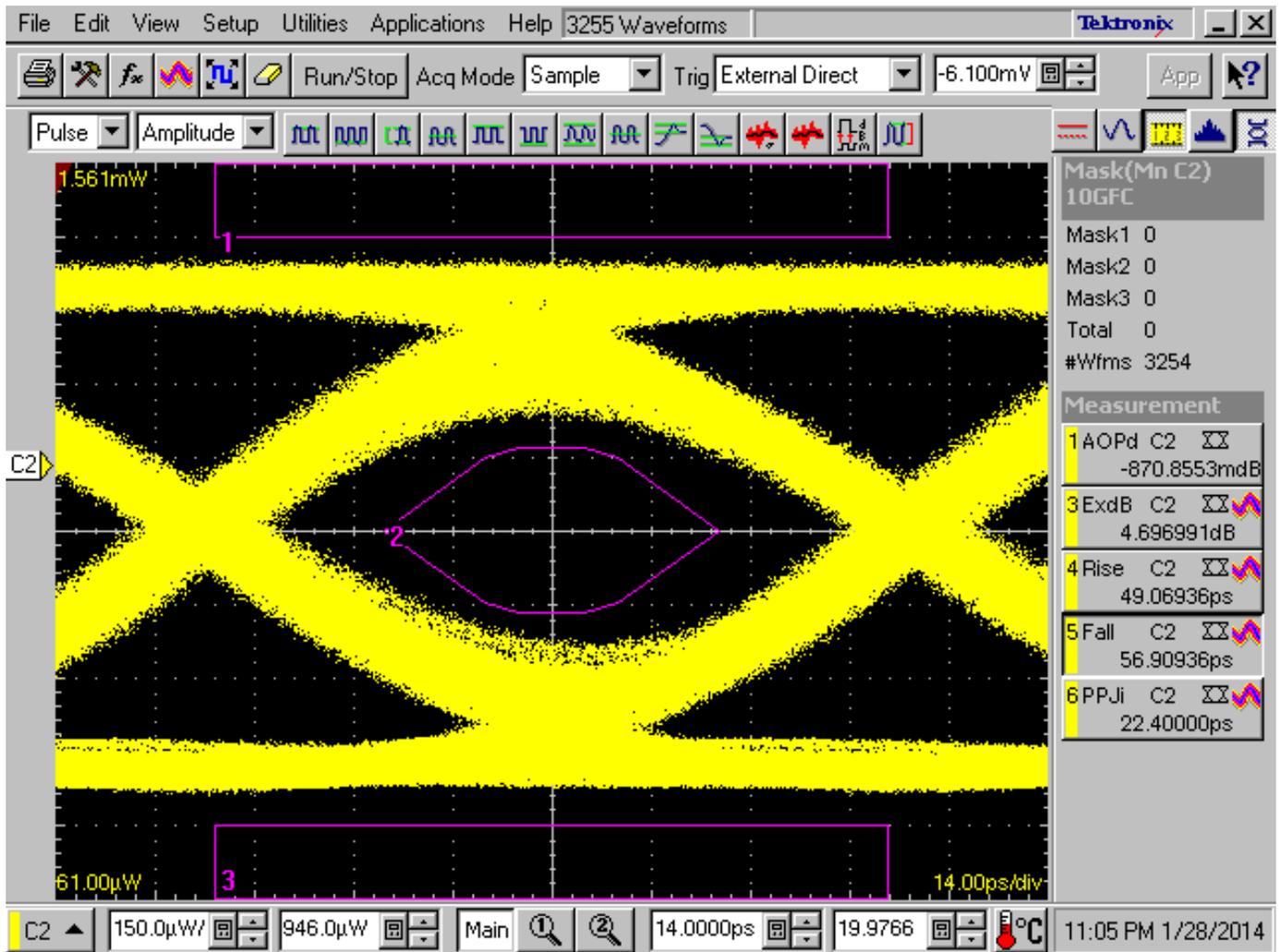

Fig. 7. Eye diagram of the MTx prototype.

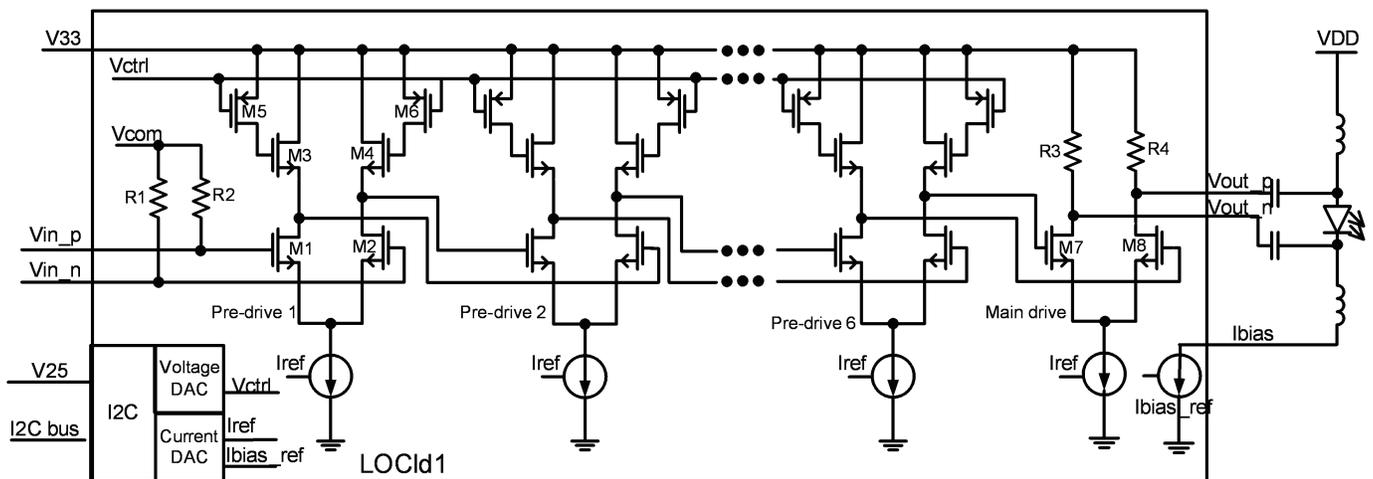

Fig. 8. Block diagram of LOCld1.

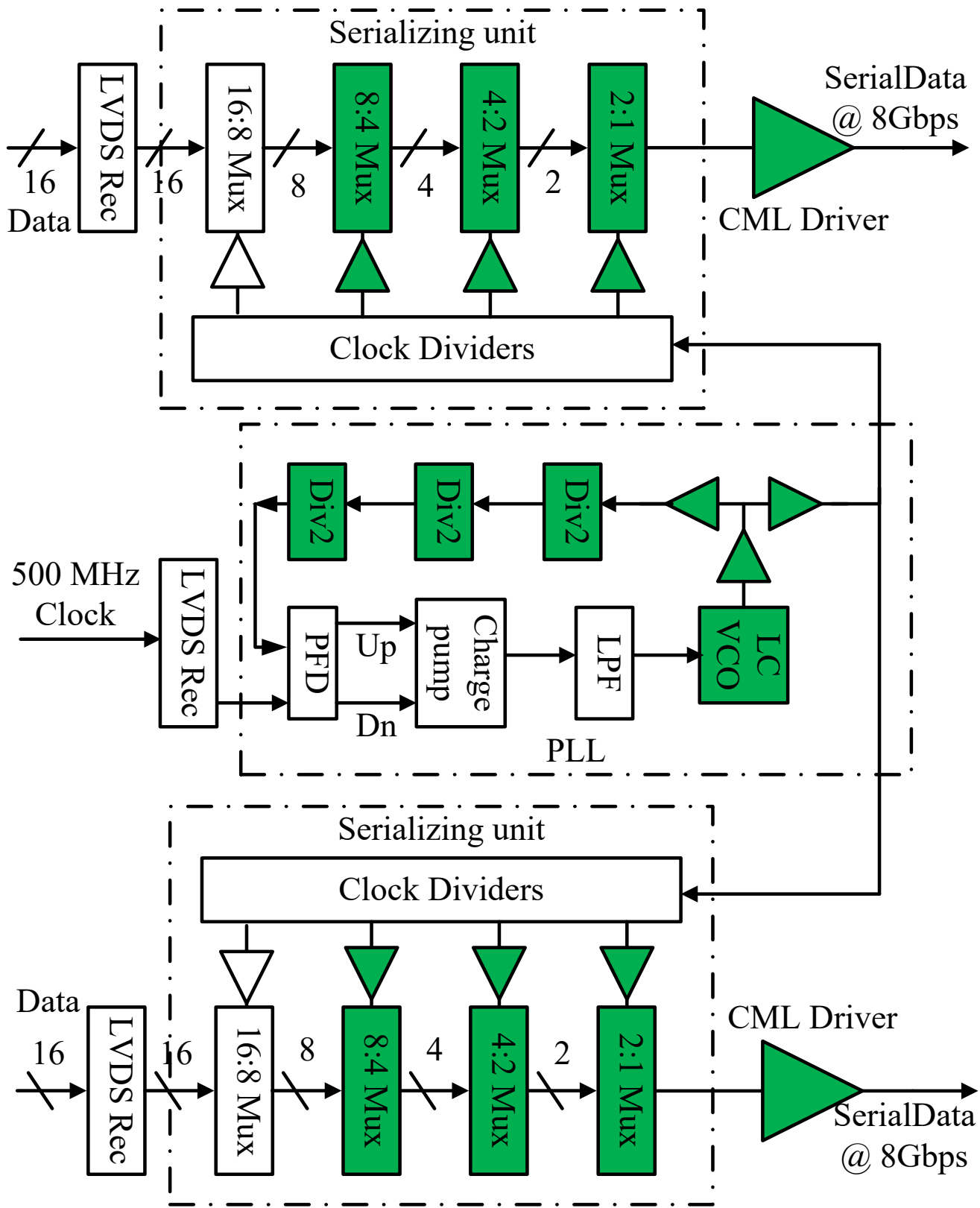

Fig. 9. Block diagram of LOCs2.

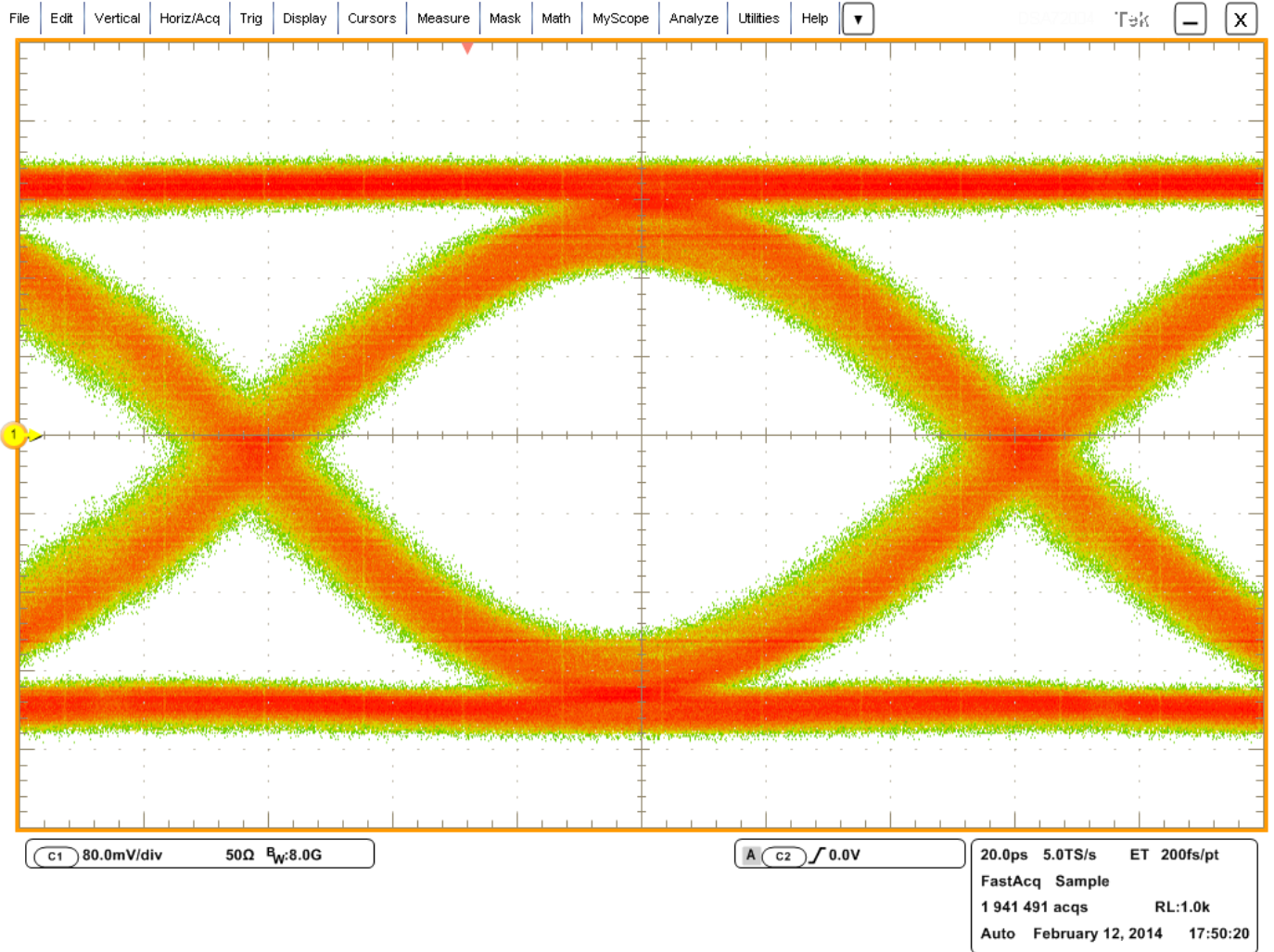

Fig. 10. Eye diagram of LOCs2 at 8 Gbps.

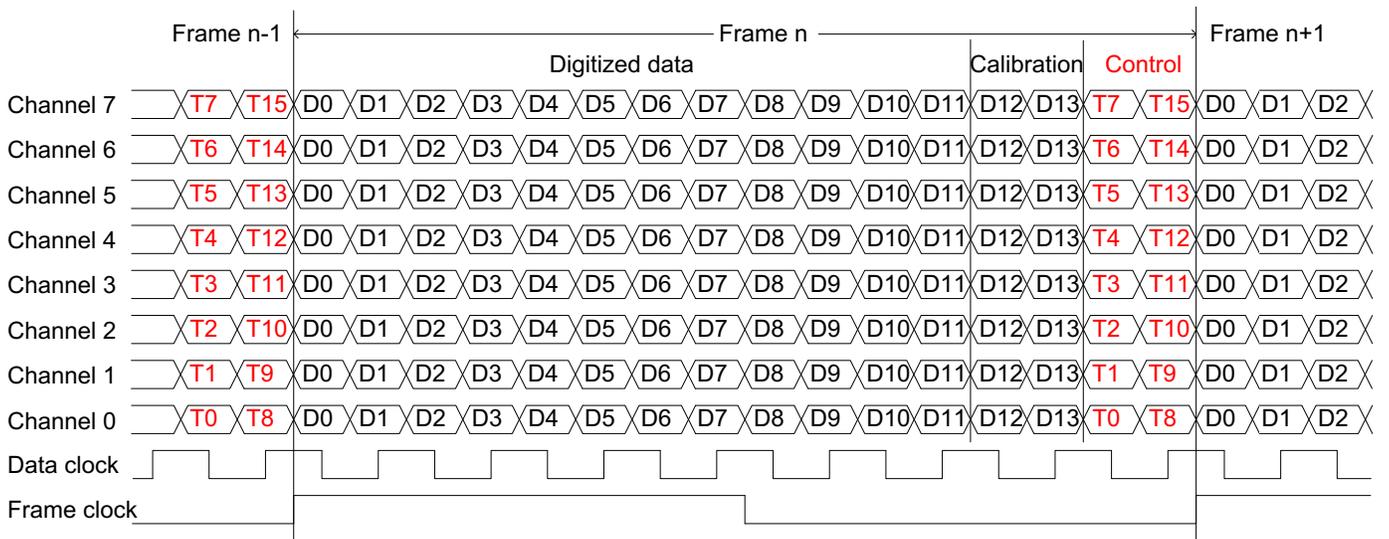

Fig. 11. Frame definition of LOCic.

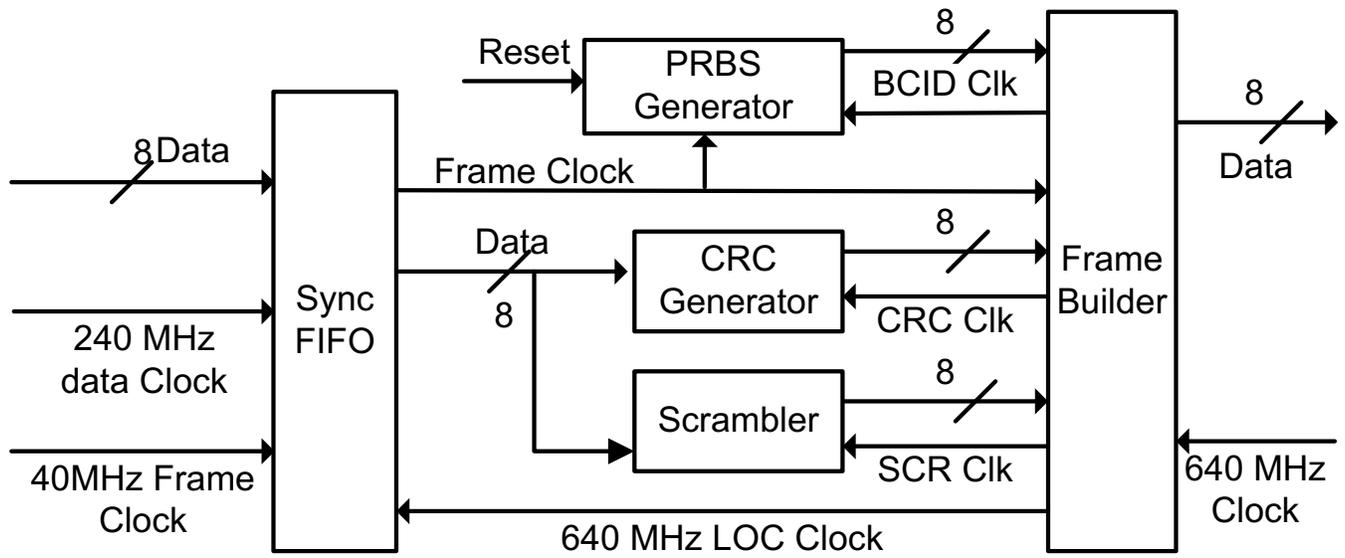
Fig. 12. Block diagram of the encoder.

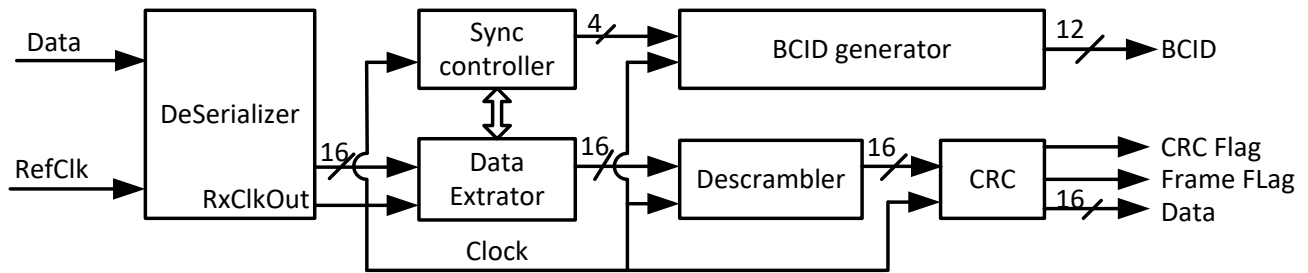
Fig. 13. Block diagram of an FPGA receiver implementation.